\documentclass[12pt]{article}
\usepackage{epsfig,amssymb}

\newcommand{\bq}{\begin{equation}}
\newcommand{\eq}{\end{equation}}
\newcommand{\bqa}{\begin{eqnarray}}
\newcommand{\eqa}{\end{eqnarray}}
\newcommand{\ben}{\begin{enumerate}}
\newcommand{\een}{\end{enumerate}}
\newcommand{\bc}{\begin{center}}
\newcommand{\ec}{\end{center}}
\newcommand{\bqb}{\begin{eqnarray*}}
\newcommand{\eqb}{\end{eqnarray*}}

\def\eg{{\it e.g. }}

\def\sw{s_W}
\def\cw{c_W}
\def\swd{s^2_W}
\def\cwd{c^2_W}

\def\mz{m_Z}
\def\mzd{m_Z^2}

\def\L{ {\cal L }}
\def\C{ {\cal C }}
\def\tchi{\tilde \chi^0}
\def\stop{\tilde t}

\def\cbeta{{\rm c}_\beta}

\def\sbeta{{\rm s }_\beta}
\def\s2beta{{\rm s}_{2 \beta}}

\def\pr#1#2#3{ Phys. Rev. ${\bf{#1}}$:#2 (#3)}

\def\prep#1#2#3{ Phys. Rep. ${\bf{#1}}$:#2 (#3)}

\def\np#1#2#3{ Nucl. Phys. ${\bf{#1}}$:#2 (#3)}
\def\zp#1#2#3{ Z. f. Phys. ${\bf{#1}}$:#2 (#3)}
\def\epj#1#2#3{ Eur. Phys. J. ${\bf{#1}}$:#2 (#3)}

\begin{document}
\pagenumbering{arabic} \thispagestyle{empty}
\def\thefootnote{\fnsymbol{footnote}}
\setcounter{footnote}{1}

\begin{flushright}
THES-TP 2001/06\\
hep-ph/0107249 \\
July 2001\\
 \end{flushright}
\vspace{2cm}
\begin{center}
{\Large\bf A  description of
 the  neutralino observables  in terms of
projectors\footnote{Partially supported by EU contracts
HPMF-CT-1999-00363 and
  HPRN-CT-2000-00149.}.}
 \vspace{1.5cm}  \\
{\large G.J. Gounaris, C. Le Mou\"{e}l and  P.I. Porfyriadis} \\
\vspace{0.4cm}
Department of Theoretical Physics, Aristotle
University of Thessaloniki,\\ Gr-54006, Thessaloniki, Greece.\\

\vspace{2.cm}

{\bf Abstract}
\end{center}

Applying Jarlskog's treatment of the CKM matrix, to the neutralino
mass matrix in  MSSM for real soft gaugino SUSY breaking and
$\mu$-parameters, we construct explicit analytical expressions for
the four projectors
 which acting on any neutralino state project out the
mass eigenstates. Analytical expressions for the neutralino mass
eigenvalues  in terms of the various  SUSY parameters, are also
given. It is shown that these projectors and mass eigenvalues are
sufficient to describe any physical observable involving
neutralinos, to any order of perturbation theory.
As an example, the $e^-e^+ \to \tilde \chi^0_i  \tilde \chi^0_j $
cross section at tree level is given in terms of these projectors.
The expected magnitude of their various matrix elements
in plausible SUSY scenarios is also discussed.\\

\noindent
PACS number(s): 12.60.Jv

\def\thefootnote{\arabic{footnote}}
\setcounter{footnote}{0}
\clearpage

\section{Introduction}

If the Minimal Supersymmetric Extension of the
Standard Model (MSSM) is realized in nature, then some at least
of the four possible neutralinos,
 should be among the lightest new particles
to exist \cite{SUSY-rev}. This is particularly true if R-parity is
conserved, so that the lightest neutralino is absolutely stable
and probably describes the Dark Matter.

Since the four neutralinos are generally mixed, their study is
complicated, and  it thus necessitates the discovery of efficient
physically motivated ways to describe it.
At the tree level, this mixing depends on the
SUSY breaking gaugino masses  $M_1,~ M_2$, and the $\mu$
Higgs sector parameter. These parameters could be real or complex
depending on whether the new (beyond the Standard Model)
interactions contained in MSSM respect the CP-symmetry or not.
There   already exist many papers where observables related
to  neutralinos have been calculated in terms of this mixing
\cite{neutralinos1a, neutralinos1b, neutralinos2, neutralinos3}.

The main purpose of the present paper it to show that
for real $(M_1,~ M_2,~\mu)$, all possible neutralino
observables  can be described in terms of the matrix elements
of four neutralino projections operators (one for each
neutralino), and the neutralino masses. For all of them,
explicit analytical expressions are then derived.
Since these projection matrix elements and masses
are themselves physically observable; a considerable economy in
the description is thereby achieved.

In order to  state our results and establish our notation,
we first note that the neutralino mass term in the minimal SUSY
Lagrangian is given by \cite{SUSY-rev}
\bq
\L_m= ~-\frac{1}{2} \Psi^{0\tau}_L \C Y \Psi^0_L ~+~ {\rm h.c.}
~~ , ~~~~  \label{mass-term}
\eq
where $\C =i\gamma^2 \gamma^0$ in (\ref{mass-term}) is the usual
Dirac charge conjugation matrix, while the L-components of all
neutralino fields are  described in the "weak basis" by the column
vector
\bq
\Psi^0_L \equiv \left ( \matrix{\tilde B_L \cr \tilde
W^{(3)}_L \cr \tilde H_{1L}^0 \cr  \tilde H_{2L}^0 \cr } \right )
~~ . \label{Psi0L}
\eq
The mass-matrix $Y$ is of course symmetric and given by
\bq
Y=  \left ( \matrix{  M_1  & 0  & -\mz \sw \cbeta &   \mz \sw
\sbeta \cr 0  & M_2  & \mz \cw \cbeta  & -\mz \cw \sbeta \cr -\mz
\sw \cbeta  & \mz \cw \cbeta  & 0  & -\mu \cr
 \mz \sw \sbeta  &  -\mz \cw \sbeta  & -\mu  & 0 \cr}
\right ) ~~ , \label{Y-matrix}
\eq
where   $\sbeta \equiv \sin\beta$,  $\cbeta \equiv \cos \beta$.
The only neutralino fields entering   the
complete MSSM Lagrangian are the components of $\Psi^0_{\alpha L}$ in
(\ref{Psi0L}) and their hermitian conjugates.
\par

If   $(M_1, ~M_2,~ \mu )$ are real, then the symmetric
matrix $Y$ in (\ref{Y-matrix}) can  be
diagonalized through the  real orthogonal transformation $U^0$
giving
\bq
U^{0\tau} Y U^0 =\left (  \matrix{\tilde m_{\tchi_1} & 0 & 0&
0 \cr 0 & \tilde m_{\tchi_2} & 0 & 0 \cr 0 & 0 & \tilde
m_{\tchi_3} & \cr 0 & 0 & 0& \tilde m_{\tchi_4}  \cr }\right ) ~~
, ~~ \label{Yd-matrix}
\eq
where the real eigenvalues  $\tilde m_{\tchi_j}$ can be of either
sign and have been ordered so that
\bq
|\tilde m_{\tchi_1}| \leq |\tilde m_{\tchi_2}| \leq |\tilde
m_{\tchi_3}| \leq |\tilde m_{\tchi_4}| ~~ . \label{mj-ordering}
\eq
We call $\tilde m_{\tchi_j}$ the  "signed" neutralino masses,
while the physical (positive) neutralino masses $m_{\tchi_j}>0$
are related to them  by
\bq
\tilde m_{\tchi_j} =\eta_j m_{\tchi_j}~~~~ {\rm with} ~~~~~
~~~~  \eta_j=\pm 1 ~~~ ~. ~  \label{etaj}
\eq

As it is well known, the signs $\eta_j$ in (\ref{etaj}) determine
the CP-eigenvalues  of the physical neutralino fields $ \tchi_{j}~
(j=1-4)$. Their L-parts
 are related to  the "weak basis"
 L-neutralino fields $\Psi^0_{\alpha L}~(\alpha=1-4)$, by
\bq
\Psi^0_{\alpha L}= \sum_{j=1}^4 U^0_{\alpha j}\tchi_{jL}
\tilde \eta_j ~~~, \label{U0-matrix}
\eq
where we define   $(\tilde \eta_j =1~ ~ {\rm or} ~~i)$
depending on whether $ (\eta_j =1~ ~ {\rm or} ~~-1)$.
 Thus,  $ \eta_j=\tilde \eta_j^2$.

 In $U^0_{\alpha j}$ at (\ref{U0-matrix}),
the first index  $\alpha$  counts the weak basis
neutralino fields, while
the second index j refers the mass eigenstate neutralinos.
Using it, we
then remark that the projector matrix to the $j$-th neutralino
state is given by
\bq
P_j=U^0 E_j U^{0\tau} ~~, \label{Projector-a}
\eq
where   the matrix elements of the four-by-four
basic matrices  $E_j$ are
$(E_j)_{ik}\equiv \delta_{ij}\delta_{jk}$,
so that   the weak basis
matrix elements of  $P_j$
are\footnote{Here of course, there is no summation over $j$.}
\[
P_{j\alpha\beta}=U^0_{\alpha j}U^0_{\beta j}~~ .
\]
 As expected, the
usual projector relations
\bqa
&& P_j P_i =P_j~\delta_{ji}
 ~~,~~ TrP_j=1 ~~ ,~~~ P_j= P_j^\tau=P_j^\dagger  ~~,
\nonumber \\
&&  Y=\sum_{j=1}^4 \eta_j m_{\tchi_j} P_j ~~ .
\label{Projector-constraint}
\eqa
are satisfied by (\ref{Projector-a}).\par

As already stated, in
this paper we first construct  explicit
 expressions for the   projectors to
 the neutralino mass eigenstates. This is done  in analogy to
 the Jarlskog's treatment of the   CKM Matrix and it is
 presented in  Section 2 \cite{Jarlskog}. In the first version of
 these expressions, the projectors are written in terms of the
 $(M_1,~ M_2~, \mu)$ parameters and the neutralino masses and CP
 eigenvalues. But versions are  also given
  in which the projectors are  expressed in terms
  of $(M_1,~ M_2~, \mu)$ and \eg the masses of only
  the  two lightest, or even only  the very lightest neutralino.
Phenomenologically these later expressions
may be more useful in situations where only one or two of
the lightest neutralino masses are known
\cite{neutralinos2, neutralinos3}.

 In the same Section 2
 we also give an analytic solution of
the characteristic equation for $Y$, which determines the signed
masses $\tilde m_{\tchi_j}$ in terms of $(M_1,~ M_2~, \mu)$.
In principle this neutralino-mass
solution should be equivalent to the one presented in
\cite{Cairo}. Nevertheless we give it here since its form is
somewhat different, and because it is  very useful for
constructing the aforementioned projectors.

In Section 3 we then show that all neutralino propagators that can
possibly appear in MSSM, are completely expressed in terms
of the above projectors and the signed neutralino masses. This
then leads to the conclusion that all  neutralino information
contained   in any cross section involving either virtual or
external neutralinos,  is fully described in terms of the
neutralino projectors and signed neutralino masses.
As an  example, we
give the tree level formulae for the $e^-e^+ \to \tchi_i \tchi_j$
 cross section expressed this way.
In the same Section  we also discuss  the
 numerical expectations for the matrix elements of
 the neutralino projectors, in various SUSY scenarios.
  A summary of the conclusions is  given in Section 4.

\vspace{1cm}
\section{The neutralino Projectors and masses. }

The characteristic equation for the  neutralino mass matrix
generated by $Y$ in (\ref{Y-matrix}) is
\bq
x^4- Ax^3+Bx^2-Cx+D=0 ~~, \label{characteristic}
\eq
with
\bqa
A &=& M_1+M_2 ~~ ,\nonumber \\
B &=& -\mzd-\mu^2 +M_1M_2 ~~ ,
\nonumber \\
C & =& -M_1 (\mu^2+\mzd \cw^2)-M_2 (\mu^2+\mzd \sw^2)
+\mu \mzd \s2beta ~~, \nonumber \\
D &=& -M_1 M_2\mu^2 +\mu \mzd
(M_1\cw^2+M_2 \sw^2) \s2beta
 ~~ , \label{ABCD-par}
\eqa
where $\s2beta \equiv 2 \sbeta \cbeta$

There should exist four real
solutions of (\ref{characteristic}) determining the signed masses
$\tilde m_{\tchi_j}$, and thereby $\eta_j$ and $m_{\tchi_j}$;
compare (\ref{etaj}). To determine them
we first construct \underline{one}  real root
of the auxiliary cubic equation \cite{Equations}
\bqa
  && \theta^3 +a\theta +b=0 ~~, \nonumber \\
&& a  \equiv  -\frac{1}{4}\left (-\frac{AC}{4} +\frac{B^2}{12}+D
\right ) ~~, \nonumber \\
&& b \equiv \frac{1}{4} \left (
-\frac{A^2 D}{16}+\frac{A B C}{48}-\frac{B^3}{216}+ \frac{B
D}{6}-\frac{C^2}{16} \right ) ~~.  \label{auxiliary-eq}
\eqa
Depending on the signs of $a$ and $\Delta$  defined as
\bq
\Delta  \equiv
\frac{b^2}{4}+\frac{a^3}{27} ~~,  \label{auxiliary-Delta}
\eq
the needed single
real root of (\ref{auxiliary-eq})   is constructed as:
\begin{itemize}
\item  if ~~ $\Delta \leq 0 ~,~a<0  $,~~ then
\bq
\cos(3\phi) \equiv -\frac{b}{2} \left (\frac{3}{|a|} \right
)^{3/2}  , ~~ \theta = 2\sqrt{\frac{|a|}{3}}\cos \left(\phi
+\frac{2n\pi}{3}\right ) \label{auxiliary1}
\eq
where any of the three choices $n=1,2,3$ may be used.
\vspace{0.5cm}
\item if ~~ $\Delta >0 ~,~a<0$, ~~ then
\bq
\cosh(3\phi)=\frac{|b|}{2}\left (\frac{3}{|a|} \right
)^{3/2}~~,~~ \theta =- 2~ {\rm  Sign}(b)
\sqrt{\frac{|a|}{3}}\cosh (\phi) ~~ , \label{auxiliary2}
\eq
\vspace{0.5cm}
\item if ~~ $\Delta >0 ~,~a>0$, ~~ then
\bq
\sinh(3\phi)=-\frac{b}{2}\left (\frac{3}{|a|} \right
)^{3/2}~~,~~ \theta = 2  \sqrt{\frac{|a|}{3}}\sinh (\phi) ~~ .
\label{auxiliary3}
\eq
\vspace{0.5cm}
\end{itemize}

We should remark at this point,
that in all MSSM case studies we are aware of, the situation
$(\Delta \leq 0 ~,~a<0)$ is met,
which indicates that (\ref{auxiliary1})
is probably the most useful case.\par

Using $\theta$  and defining also
\bq
E=\frac{1}{4} \left (A^2-\frac{8B}{3}+16\theta \right)^{1/2}
~~~, ~~~ F=\frac{1}{4E}\left (C-\frac{A B}{6}-2 A\theta \right )
~~ ,
\eq
we obtain the four signed neutralino masses from
\bqa
\tilde m_{\tchi_j}\equiv \eta_j m_{\tchi_j}
&= & \frac{1}{2} \Big \{ \Big (\frac{A}{2}-2 E\Big )
\pm \sqrt{\Big (\frac{A}{2}-2 E\Big )^2-
4 \Big (\frac{B}{6}+2 \theta +F\Big )}
\Big \} ~~ , \nonumber \\
&=& \frac{1}{2} \Big \{ \Big (\frac{A}{2} +2 E\Big )
\pm \sqrt{\Big (\frac{A}{2}+2 E\Big )^2- 4
\Big (\frac{B}{6}+2 \theta - F\Big )} \Big \} ~~ .
\label{tmchij}
\eqa

The general expressions for  the parameters $(A-F)$ do not allow us
to order equations  (\ref{tmchij}), so that
(\ref{mj-ordering}) is satisfied in any model.
This can efficiently be done in specific
models though. As such we choose to consider the thirteen
 benchmark SUGRA scenarios  of
\cite{Ellis-bench}, which are more or less consistent
with what is presently known
from  LEP, $b\to s \gamma $, $g_\mu-2$ and
cosmology.
As the scale in which to apply relations (\ref{tmchij})
(as well as the formulae for the neutralino projectors below)
 we take\footnote{Here $m_{\stop_1}$
  and $m_{\stop_2}$ are the masses of the two stop quarks.}
 $Q=\sqrt{m_{\stop_1}  m_{\stop_2}}$,  in agreement with the one chosen
in Table 3 of \cite{Ellis-bench}. This allows us to
check   that in all cases
the absolute values of $\tilde m_{\tchi_1}$,    $\tilde m_{\tchi_2}$,
$\tilde m_{\tchi_3}$,  $\tilde m_{\tchi_4}$ agree with those
quoted there. For all the above scenarios
we then find that  $\tilde m_{\tchi_3}$  is negative, while all other
 ``signed'' masses are positive.
In Table 1 we  present them together with the
  $(M_1,~ M_2,~ \mu)$-parameters at the scale
$Q$, for  seven of the benchmark scenarios of \cite{Ellis-bench},
which are   characterised  by neutralinos that  are  sufficiently light
to be producible in the future Colliders.
At the end of the next Section, we come back to the
consideration of these  scenarios.
\begin{table}[htb]
\begin{center}
{ Table 1: Parameters at the scale $Q=\sqrt{m_{\stop_1}
m_{\stop_2}}$, for some SUGRA
 scenarios of \cite{Ellis-bench}.
  (Dimensions in $\rm GeV$.)}\\
  \vspace*{0.3cm}
\begin{tabular}{||c|c|c|c|c|c|c|c|c|c||}
\hline \hline Scenario& Q & $M_2$ & $M_1$ & $\mu$ & $\tan \beta$ &
$\tilde m_{\tchi_1}$ &  $\tilde m_{\tchi_2}$ & $\tilde
m_{\tchi_3}$ &  $\tilde m_{\tchi_4}$   \\ \hline
 A & 1050 & 480  &254 & 768 & 5 & 252 & 467 &-770 & 785  \\
 C &726 & 317 & 165 & 520 & 10 & 163 & 303 & -524 & 540 \\
 D &937 & 421 & 221 & -662 & 10 & 221 & 414 & -667 &674 \\
 E &1129 & 245 & 125 & 255 & 10 & 117 & 197 & -262 & 318 \\
 G &687 &  299 & 161 & 485 & 20 & 159 & 286 &
-490 & 505  \\
 I &650 &  279 & 146 & 454 & 35 & 144 & 266 & -460 & 475\\
 L &826 & 361 & 187 & 560 & 45 & 186 & 349& -565 &578 \\
 \hline \hline
\end{tabular}
\end{center}
\end{table}

\vspace{0.5cm}
We next turn to the  projectors to  the neutralino $\tchi_j$
mass-eigenstate, defined in (\ref{Projector-a}).
In the most interesting case that all
neutralino masses are different from each other,
these projectors may be written,  in analogy to
the CKM treatment of Jarlskog \cite{Jarlskog},
as\footnote{If needed, the generalization to the improbable
case of neutralino mass degeneracy can
easily be done following the instructions  in \cite{Jarlskog}.}

\bqa
P_1 &=& \frac{(\tilde m_{\tchi_4} -Y) (\tilde m_{\tchi_3}
-Y)(\tilde m_{\tchi_2} -Y)} {(\tilde m_{\tchi_4} -\tilde
m_{\tchi_1}) (\tilde m_{\tchi_3} -\tilde m_{\tchi_1}) (\tilde
m_{\tchi_2} -\tilde m_{\tchi_1})} ~~ ~~, \nonumber \\[0.3cm]
P_2 &=& \frac{(\tilde m_{\tchi_4} -Y) (\tilde m_{\tchi_3} -Y)
(Y- \tilde m_{\tchi_1} )} {(\tilde m_{\tchi_4} -\tilde m_{\tchi_2})
(\tilde m_{\tchi_3} -\tilde m_{\tchi_2}) (\tilde m_{\tchi_2}
-\tilde m_{\tchi_1})} ~~ ~~, \nonumber \\[0.3cm]
P_3 &=&
\frac{(\tilde m_{\tchi_4} -Y) (Y- \tilde m_{\tchi_2} )(Y- \tilde
m_{\tchi_1} )} {(\tilde m_{\tchi_4} -\tilde m_{\tchi_3}) (\tilde
m_{\tchi_3} -\tilde m_{\tchi_2}) (\tilde m_{\tchi_3} -\tilde
m_{\tchi_1})} ~~ ~~, \nonumber \\[0.3cm]
P_4 &=& \frac{(Y- \tilde
m_{\tchi_3} ) (Y- \tilde m_{\tchi_2} )(Y- \tilde m_{\tchi_1} )}
{(\tilde m_{\tchi_4} -\tilde m_{\tchi_3}) (\tilde m_{\tchi_4}
-\tilde m_{\tchi_2}) (\tilde m_{\tchi_4} -\tilde m_{\tchi_1})} ~~
~~, \label{Projector-b}
\eqa
where $Y$ is given in (\ref{Y-matrix}).

In (\ref{Projector-b}), the  projectors are written in terms of
the matrix $Y$ and all the "signed" neutralino masses. By using the
explicit form of $Y$, and the theory of characteristic
equations \cite{Matrices},
a more useful  expression is obtained,  which, apart from
$Y$-matrix elements, involves
only the corresponding signed neutralino mass.
To present them we define $(i=1-4)$,
\bq
P_i=\frac{\tilde{P}_i}{\Delta_i} ~~, \label{Projector-c}
\eq
with
\bq
 \Delta_i \equiv -3 \tilde m_{\tchi_i}^4+2 \tilde m_{\tchi_i}^3
A - \tilde m_{\tchi_i}^2 B+D ~~ ,\label{Delta-i}
 \eq
where $A,B,D$ are  given  in (\ref{ABCD-par}).
 The matrix elements for $\tilde{P}_i$ $(i=1,4)$ then read
\bqa
&& \tilde{P}_{i11}=-\tilde m_{\tchi_i}^3 M_1+\tilde
m_{\tchi_i}^2
 (M_1 M_2-\mzd \swd)+\tilde m_{\tchi_i}[M_1 \mu^2+
 \mzd (M-\mu \swd \s2beta)] +D ~, \nonumber  \\
&& \tilde{P}_{i22}=-\tilde m_{\tchi_i}^3 M_2+\tilde m_{\tchi_i}^2
 (M_1 M_2-\mzd \cwd)+\tilde m_{\tchi_i}[M_2 \mu^2+\mzd (M-\mu \cwd
 \s2beta)] +D~ , \nonumber \\
&& \tilde{P}_{i33}=-\tilde m_{\tchi_i}^2 (\mu^2+\mzd \cbeta^2)
  + \tilde m_{\tchi_i} [\mu^2 (M_1+M_2)+\mzd (-\mu \
  \s2beta+M \cbeta^2 )]+D ~, \nonumber  \\
&&  \tilde{P}_{i44}=-\tilde m_{\tchi_i}^2 (\mu^2+\mzd \sbeta^2)
  +\tilde m_{\tchi_i} [\mu^2 (M_1+M_2)+\mzd (-\mu \
  \s2beta+M \sbeta^2 )]+D ~ , \nonumber \\
&& \tilde{P}_{i12}= \tilde{P}_{i21}=\tilde m_{\tchi_i} \mzd \sw
\cw (\tilde  m_{\tchi_i}+\mu \ \s2beta)~~, \nonumber \\
&& \tilde{P}_{i13}= \tilde{P}_{i31}=\tilde m_{\tchi_i} \mz \sw
(\tilde  m_{\tchi_i}-M_2) (\tilde m_{\tchi_i} \cbeta+\mu
\ \sbeta) ~~, \nonumber \\
&& \tilde{P}_{i14}= \tilde{P}_{i41}=- \tilde m_{\tchi_i} \mz \sw
(\tilde  m_{\tchi_i}-M_2) (\tilde m_{\tchi_i} \sbeta
+\mu \ \cbeta) ~, \nonumber \\
&& \tilde{P}_{i23}= \tilde{P}_{i32}=- \tilde m_{\tchi_i} \mz \cw
(\tilde  m_{\tchi_i}-M_1) (\tilde m_{\tchi_i} \cbeta
+\mu \ \sbeta) ~~, \nonumber  \\
&&  \tilde{P}_{i24}= \tilde{P}_{i42}= \tilde m_{\tchi_i} \mz \cw
(\tilde  m_{\tchi_i}-M_1) (\tilde m_{\tchi_i} \sbeta
+\mu \ \cbeta) ~, \nonumber \\
&& \tilde{P}_{i34}= \tilde{P}_{i43}
\nonumber \\
&& ~~~~~~~~ =\tilde m_{\tchi_i} [\tilde
m_{\tchi_i}^2 \mu+  \tilde m_{\tchi_i} [-\mu (M_1+M_2)
+\mzd \frac{\s2beta}{2}]
  +\mu M_1 M_2 -M \mzd \frac{\s2beta}{2}] ~~ ,
\label{tProjector}
 \eqa
where $M \equiv M_1 \cwd+M_2 \swd$. As seen from
(\ref{Delta-i},   \ref{tProjector}),
each of these  projector expression only involves
the corresponding neutralino
mass $\tilde m_{\tchi_i}$ and the
$(M_1,~ M_2, ~\mu)$-parameters. We have verified that
these expressions satisfy the constraints
(\ref{Projector-constraint}).  \par

As advocated in \cite{neutralinos2, neutralinos3}, when the
neutralinos will start being discovered, it would be
interesting to consider situations where only $\tchi_1$,
or $\tchi_1$ and $\tchi_2$, have been seen.
In such a case it would be advantageous to have projector
expressions, in which only  the lightest one or two
neutralino masses will be contained.\par

Thus if we assume that $m_{\tchi_1},~ m_{\tchi_2}$ as well as the
the signs $\eta_1, \eta_2$ are already known, then the expressions
(\ref{ABCD-par}) for $A, B, C, D$ and their connection  to
the $Y$-eigenvalues determine\footnote{Notice that the requirement
$|\tilde m_{\tchi_3}|\leq |\tilde m_{\tchi_4}|$, fully determines the
identification of the solutions.} \cite{Matrices}
\bq
\tilde m_{\tchi_3}~,~ \tilde m_{\tchi_4}=
\frac{1}{2}\Bigg \{
A- \tilde m_{\tchi_1}-\tilde m_{\tchi_2} \pm
\sqrt{(A- \tilde m_{\tchi_1}-\tilde m_{\tchi_2})^2
-\frac{4 D}{\tilde m_{\tchi_1}\tilde m_{\tchi_2}} }
\Bigg \} ~, \label{mchi34}
\eq
so that the projectors in
(\ref{Projector-c}, \ref{tProjector}) can be expressed in
terms of just the signed masses of the
 two lightest neutralinos, and an independent combination of
 $M_1,~M_2,~\mu$ \cite{neutralinos2}.\par

 Furthermore,  by using the
equations in Appendix B of \cite{neutralinos2}, we can
also express all projector elements in
(\ref{Projector-c},\ref{tProjector}) in terms of just
$\tilde m_{\tchi_1}$ and the needed two independent
combinations of the
relevant SUSY parameters. Thus, the projector formalism can be
easily adapted to the various situations that might appear during
the neutralino  searching, \cite{neutralinos2, neutralinos3}.\par

 To avoid any possible confusion, we should  emphasize at this point
that the above formalism for  the neutralino masses and projectors is
 mathematically (and numerically)  equivalent to the approach  of \cite{Cairo}.

\section{The neutralino Observables in MSSM.}

Since the only neutralino fields entering the  MSSM Lagrangian,
are the "weak-basis" fields  appearing in  (\ref{Psi0L}) and their
hermitian conjugates;  it follows that the only
neutralino propagators that
can contribute in any possible diagram are just
\bq
 \langle 0|\Psi_{\alpha L}^0(x)\Psi_{\beta L}^{0\tau}(y)
|0 \rangle ~~~ , ~~~
\langle 0|\Psi_{\alpha L}^{0\dagger \tau}(x)
\Psi_{\beta L}^{0 \dagger}(y)
|0 \rangle ~~~ , ~~~
\langle 0|\Psi_{\alpha L}^{0}(x)\Psi_{\beta L}^{0 \dagger}(y)
|0 \rangle ~~~ ,
\label{neutralino-prop}
\eq
where $(\alpha, \beta)$ count the weak-basis fields.

Denoting then the elementary propagator functions as
\bqa
\Delta_F(x-y;~ m) &= & i \int \frac{d^4k}{(2\pi)^4}
e^{-i k(x-y)}\frac{1}{k^2-m^2+i\epsilon} ~~ ,
\nonumber \\
S_F^{(1)}(x-y; ~ m) &= & i \int \frac{d^4k}{(2\pi)^4}
e^{-i k(x-y)}\frac{\rlap /k }{k^2-m^2+i\epsilon} ~~ ,
\label{propagator-functions}
\eqa
we find that the weak-basis neutralino propagators of
(\ref{neutralino-prop}) are given at lowest order by
\bqa
&& \langle 0|\Psi_{\alpha L}^0(x)\Psi_{\beta L}^{0\tau}(y)
|0 \rangle = -~
\langle 0|\Psi_{\alpha L}^{0\dagger \tau}(x)
\Psi_{\beta L}^{0 \dagger}(y)
|0 \rangle \nonumber \\
&& = -\sum_{j=1}^4 \eta_j m_{\tchi_j}P_{j\alpha \beta}
\Delta_F(x-y;~ m_{\tchi_j}) ~\C  ~\frac{(1-\gamma_5)}{2}
 ~~ , \label{neutralino-prop1} \\
&& \langle 0|\Psi_{\alpha L}^{0}(x)\Psi_{\beta L}^{0 \dagger}(y)
|0 \rangle = \sum_{j=1}^4 P_{j\alpha \beta}
S_F^{(1)}(x-y; ~ m_{\tchi_j} ) \gamma_0 ~\frac{(1-\gamma_5)}{2}
 ~~ , \label{neutralino-prop2}
\eqa
where the usual charge conjugation Dirac matrix
 $\C$ has been given  immediately after
(\ref{mass-term}). Notice that (\ref{neutralino-prop1}) only
involves the scalar propagator function in
(\ref{propagator-functions}).
As it is seen from (\ref{neutralino-prop1}, \ref{neutralino-prop2}),
 the $\eta_j$ signs always appear multiplied by the
corresponding neutralino masses.

Therefore, any neutralino exchange contribution in any SUSY diagram,
is  fully described by the neutralino signed masses and
the projectors we have already constructed.
The Cutkosky rules then  guarantee, that the
observable contribution from any external
neutralino  can also always be described in terms of
its signed mass and projector. Therefore, the signed masses and
projectors contain all observable neutralino contributions
to any order of perturbation theory.\par

\vspace{0.5cm}
As an example of this result, we give below the tree level
differential cross
section for $e^- e^+ \to \tchi_i \tchi_j$. The contributions to this
process  arise from $s$-channel $Z$-exchange, and $t$-
and $u$-channel $\tilde e_L$ and $\tilde e_R$ exchanges.
The  differential cross
section may then be written as \cite{neutralinos1a}
\bq
\frac{{\rm d} \sigma (e^- e^+ \to \tchi_i \tchi_j)}{{\rm d}
t}= \frac{\alpha^2 \pi}{4 s^2(1+\delta_{ij})}
\Big [\Sigma_Z+\Sigma_{\tilde e_L}+
\Sigma_{\tilde e_R} +\Sigma_{Z \tilde e_L}+\Sigma_{\tilde e_R}
 \Big ] ~, \label{chiij-cross}
\eq
with the r.h.s. representing  respectively  the $Z$-,
$\tilde e_L$- and $\tilde e_R$-square contributions, as well as the
$Z\tilde e_L$- and $Z\tilde
e_R$-interferences\footnote{To the extend that we neglect the
electron mass, there is never any
$\tilde e_L \tilde e_R$-interference;
neither any Higgsino-$e\tilde e$ coupling.}.
These can be written as
\bqa
&& \Sigma_Z =
\frac{g_{ve}^2+g_{ae}^2}{2\sw^4\cw^4 (s-\mzd )^2}
[P_{i33}P_{j33}+P_{i44}P_{j44}-2P_{i34}P_{j34}]
\nonumber \\
& & ~~~~~~ \cdot  [(t-m_{\tchi_i}^2)(t-m_{\tchi_j}^2)+
(u-m_{\tchi_i}^2)(u-m_{\tchi_j}^2)-
2 s \eta_i \eta_j m_{\tchi_i} m_{\tchi_j}] ~ ,
\label{Z-square} \\
&& \Sigma_{\tilde e_L}= \frac{1}{4 \sw^4 \cw^4}
[\cwd P_{i22}+\swd P_{i11}+2\sw\cw P_{i12} ]
[\cwd P_{j22}+\swd P_{j11}+2\sw\cw P_{j12} ]
\nonumber \\
&& ~~~~~~ \cdot \Bigg  [\frac{(t-m_{\tchi_i}^2)(t-m_{\tchi_j}^2)}
{(t-m_{\tilde e_L}^2)^2} +
\frac{(u-m_{\tchi_i}^2)(u-m_{\tchi_j}^2)}
{(u-m_{\tilde e_L}^2)^2}
- ~\frac{2 s \eta_i \eta_j m_{\tchi_i} m_{\tchi_j}}
{(t-m_{\tilde e_L}^2)(u-m_{\tilde e_L}^2)} \Bigg ]
~ , \label{seL-square} \\
&& \Sigma_{\tilde e_R}=
\frac{4 P_{i11}P_{j11}}{ \cw^4}
 ~ \Bigg  [\frac{(t-m_{\tchi_i}^2)(t-m_{\tchi_j}^2)}
{(t-m_{\tilde e_R}^2)^2} +
\frac{(u-m_{\tchi_i}^2)(u-m_{\tchi_j}^2)}
{(u-m_{\tilde e_R}^2)^2} \nonumber \\
&&~~~~~~ - ~\frac{2 s \eta_i \eta_j m_{\tchi_i} m_{\tchi_j}}
{(t-m_{\tilde e_R}^2)(u-m_{\tilde e_R}^2)} \Bigg ]
~ , \label{seR-square} \\
&& \Sigma_{Z \tilde e_L} =
-~ \frac{ (g_{ve}+g_{ae})}{2  \sw^4\cw^4 (s-\mzd)}
\Big [\cwd (P_{i23}P_{j23}-P_{i24}P_{j24})+
\swd (P_{i13}P_{j13}-P_{i14}P_{j14}) \nonumber \\
&&  + \cw\sw (P_{i23}P_{j13}+P_{i13}P_{j23}
-P_{i24}P_{j14}-P_{i14}P_{j24}) \Big ]
\nonumber \\
&&   \cdot \Bigg  [\frac{(t-m_{\tchi_i}^2)(t-m_{\tchi_j}^2)
- s \eta_i \eta_j m_{\tchi_i} m_{\tchi_j}}{(t-m_{\tilde e_L}^2 )}
+\frac{(u-m_{\tchi_i}^2)(u-m_{\tchi_j}^2)
- s \eta_i \eta_j m_{\tchi_i} m_{\tchi_j}}{(u-m_{\tilde e_L}^2 )}
\Bigg ]  , \label{ZseL-int} \\
&& \Sigma_{Z \tilde e_R} =
 \frac{ 2 (g_{ve}-g_{ae})}{ \sw^2\cw^4 (s-\mzd)}~
[P_{i13}P_{j13}-P_{i14}P_{j14}]
\nonumber \\
&&   \cdot \Bigg  [\frac{(t-m_{\tchi_i}^2)(t-m_{\tchi_j}^2)
- s \eta_i \eta_j m_{\tchi_i} m_{\tchi_j}}{(t-m_{\tilde e_R}^2 )}
+\frac{(u-m_{\tchi_i}^2)(u-m_{\tchi_j}^2)
- s \eta_i \eta_j m_{\tchi_i} m_{\tchi_j}}{(u-m_{\tilde e_R}^2 )}
\Bigg ]  , \label{ZseR-int}
\eqa
where $g_{ve}=-0.5+2\swd$ and $g_{ae}=-0.5$ are the vector and
axial $Zee$-couplings.

These results have already been presented in \cite{neutralinos1a}.
The only new  thing here is that
they have  been expressed in terms of the neutralino
projectors. Notice that, since the
neutral gauginos  do not couple to the gauge bosons in MSSM,
the $Z$-square contribution in (\ref{Z-square}) only
depends on the Higgsino-Higgsino matrix elements of the
$\tchi_i$ and $\tchi_j$  projectors. Similarly, the
$\tilde e_R$-square contribution in (\ref{seR-square}) only
depends on the Bino-Bino elements of the same projectors.
For the $\tilde e_L$-square contribution in (\ref{seL-square})
though, the non-vanishing of both the Bino and Wino couplings
allow the appearance of the Bino-Bino, Wino-Wino and Bino-Wino
matrix elements  of   $P_i$ and $P_j$.
The $Z\tilde e_R$-interference only depends on
the Bino-Higgsino matrix elements of $P_i,~ P_j$;
while in the  $Z\tilde e_L$-interference case,
 the  Wino-Higgsino matrix
elements of the same projectors also appear;
compare (\ref{ZseL-int}, \ref{ZseR-int}).\par

\vspace{0.5cm}
We next turn to the numerical expectations for the various matrix
elements of the neutralino projectors. These can easily be
constructed in any model with real $(M_1,~ M_2, ~ \mu)$
parameters, by using either the set of equations
 (\ref{Projector-b}),  or the set
(\ref{Projector-c}, \ref{tProjector}) and the results
of Section 2.
An obvious remark we  should nevertheless
probably make, is that each such set of  equations has to be used at
a definite scale. This  means that in case two neutralinos have
very different masses, we would have to use the appropriately
scaled different values for the $(M_1,~ M_2, ~ \mu)$ parameters,
 in order to determine their  projectors  at their physical masses.

As an example of such projectors, we turn again to
the supergravity inspired
benchmark scenarios  of\footnote{Other related work may be
found in \cite{other}.}  \cite{Ellis-bench}.
For the same scenarios  as in Table 1,
we  give in  Table 2 the implied  matrix elements of the
four neutralino projectors.
In these scenarios $\tan\beta$ is varying between 5 and 45.
We note that the results in Tables 1 and  2
are at the same scale as the corresponding
  benchmark parameters appearing
in  Table 3  of \cite{Ellis-bench}, and may be combined
for studying various SUSY appearances\footnote{This of course implies
that an appropriate adjustment of all these parameters
 to the scale of the specific physical process
considered, may be needed.}.

{\small
\begin{table}[htb]
\begin{center}
{ Table 2: The Matrix elements of   the neutralino
projectors at the scale $Q=\sqrt{m_{\stop_1} m_{\stop_2}}$,
for  the SUGRA scenarios of  Table 1,
\cite{Ellis-bench}.}\\
\vspace*{0.3cm}
\begin{small}
\begin{tabular}{||c|c|c|c|c|c|c|c|c|c|c|c||}
\hline \hline  & $\tchi_j$ & $P_{j11}$ &  $P_{j22}$ & $P_{j33}$ &
$P_{j44}$ &  $P_{j12}$ & $P_{j13}$ & $P_{j14}$ & $P_{j23}$ &
$P_{j24}$ & $P_{j34}$   \\ \hline
 A    & $\tchi_1$ &0.994 & 0 &0.005  &
0.001 &-0.016 &0.068 &-0.034 &-0.001 &0.001 &-0.002  \\
   & $\tchi_2$  &0.001 & 0.953 &0.03  &0.016
&0.032  & -0.006 &0.004 &-0.17 &0.123 &-0.022  \\
   &$\tchi_3$   &0.001 &0.001  &0.497  & 0.501
&-0.001  &-0.017 &-0.017 &0.025 &0.025 &0.499  \\
   & $\tchi_4$  &0.005 & 0.046 &0.468  &0.482
&-0.014  &-0.046 &0.047 &0.146 &-0.148 &-0.475  \\
  \hline
 C   & $\tchi_1$  &0.988 &0.001  &0.01  &0.001 &-0.025  &0.099
&-0.04 &-0.003 & 0.001&-0.004  \\
   & $\tchi_2$  &0.003 &0.923  &0.052  & 0.022
&0.052  &-0.012 &0.008 &-0.219 &0.141 & -0.034 \\
   &$\tchi_3$   &0.002 &0.004  &0.492  &0.502
&-0.002  &-0.028 &-0.029 &0.042 &0.043 &0.497  \\
   & $\tchi_4$  &0.008 &0.072  &0.445  &0.474
&-0.024  &-0.059 &0.061  &0.18 &-0.185 &-0.46  \\ \hline
 D   & $\tchi_1$  &0.994 &0  &0.005  &0 &-0.004
&-0.072 &-0.017 &0 &0 &0.001  \\
   & $\tchi_2$  &0 & 0.957 & 0.032 &0.01
&0.018  &0.003 &0.002 &0.175 &0.098 &0.018  \\
   &$\tchi_3$   &0.001  &0.003  &0.495  &0.501
&-0.002  &0.027 &-0.027 &-0.04 &0.040 &-0.498  \\
   & $\tchi_4$  &0.004 &0.039  &0.468  &0.489
& -0.012 & 0.042  & 0.043 &-0.136 &-0.139 &0.478   \\
  \hline
 E    & $\tchi_1$  &0.903 &0.010  &0.067  &0.019 &-0.098  &0.246
&-0.132 &-0.026 &0.014 &-0.036  \\
   & $\tchi_2$  &0.071 &0.536  &0.237  &0.155
&0.197  &-0.13 &0.105 &-0.357 &0.288 &-0.192  \\
   &$\tchi_3$   &0.005 &0.010  &0.476  &0.509
&-0.007  &-0.05 &-0.053 &0.07 &0.071 &0.492  \\
   & $\tchi_4$  &0.019 &0.444  &0.221  &0.316
&-0.091  &-0.065 &0.077 &0.313 &-0.375 &-0.265  \\
  \hline
 G   & $\tchi_1$  &0.986 &0.001  &0.011  &0.002 &-0.026  &0.106
&-0.04 &-0.003 &0.001 &-0.004  \\
   & $\tchi_2$  &0.003 &0.917  &0.058  &0.022
&0.054  &-0.014 &0.008 &-0.23 &0.143 &-0.036  \\
   &$\tchi_3$   &0.002 &0.005  &0.491  &0.502
& -0.003& -0.032  &-0.032 &0.048 &0.048 & 0.497  \\
   & $\tchi_4$  &0.008 &0.078  &0.44  &0.474
&-0.026  &-0.061 &0.063 &0.185 &-0.192 &-0.457  \\
  \hline
 I   & $\tchi_1$  &0.985 &0.001  &0.013  &0.001
&-0.025  &0.112 &-0.038 &-0.003 &0.001 &-0.004  \\
   & $\tchi_2$  &0.003 &0.911  &0.063  &0.023
&0.056  &-0.015 &0.009 &-0.24 &0.145 &-0.038  \\
   &$\tchi_3$   &0.002 &0.006  &0.49  &0.502
&-0.004  &-0.035 &-0.035 &0.052 &0.053 &0.496  \\
   & $\tchi_4$  &0.009 &0.083  &0.435  &0.473
&-0.027  &-0.062 &0.065 &0.19 &-0.198 &-0.454  \\
  \hline
 L   & $\tchi_1$  &0.991 &0  &0.008  &0.001 &-0.015
 &0.09 &-0.032 &-0.001 &0 &-0.003  \\
   & $\tchi_2$  &0.002 &0.929  &0.05  &0.02
&0.038  &-0.009 &0.006 &-0.215 &0.137 &-0.032  \\
   &$\tchi_3$   &0.002 &0.004  &0.493  &0.502
&-0.002  &-0.028 &-0.029 &0.042 &0.042 &0.497  \\
   & $\tchi_4$  &0.006 &0.068  &0.449  &0.477
&-0.021  &-0.053 &0.055 &0.174 &-0.179 &-0.463  \\
  \hline  \hline
\end{tabular}
\end{small}
\end{center}
\end{table}
}

Following \cite{Ellis-bench}, we discuss separately below
the scenarios ( A,C,D,G,I,L) and  E respectively.
\begin{itemize}
\item  A,C,D,G,I,L.

The projector for the lightest neutralino
$\tchi_1$ in all these cases, is such that the $P_{111}$
element is much larger than all the rest. Thus, $\tchi_1$
 is always almost a pure Bino.

The  next neutralino $\tchi_2$  is dominated by the Wino,
but the admixture from the two Higgsinos is not absolutely
negligible; compare $P_{222}$ with $P_{223}$ and $P_{224}$.
In fact this admixture tends to increase with the
ordering of the scenarios appearing in
 Tables 1 and 2.

The next two neutralinos tend to consist of
roughly equal mixtures of the two Higgsinos.
In fact, from the signs
of $P_{j34}$  in  Table 2, we see that for the
  A, C, G, I and  L scenarios,
 $\tchi_3 \sim  (\tilde H_1^0 + \tilde H_2^0)/\sqrt{2} $; while
for   D,  (which is the only $\mu<0$ case)
 $\tchi_3 \sim   (\tilde H_1^0 - \tilde
H_2^0)/\sqrt{2}$. In all cases,  $\eta_3=-1$.

For $\tchi_4$, we approximately have
 $\tchi_4 \sim  (\tilde H_1^0 - \tilde H_2^0)/\sqrt{2} $
  for  A, C, G, I, L, and
  $\tchi_4 \sim   (\tilde H_1^0 + \tilde H_2^0)/\sqrt{2}$
for   D. But the  admixture from the Wino is somewhat larger
in this   case.

\item  E

This is one of the two "focus-point" scenarios in
\cite{Ellis-bench}, corresponding to  $m_0\gg m_{1/2} $ and
$\tan\beta=10$. Again $\tchi_1$ is mainly a Bino, but the
admixture from the two Higgsinos is not negligible; compare
$P_{113}$ and $P_{114}$ from Table 2.

The  most important  contribution to $\tchi_2$ comes from the Wino;
but the contributions from the Higgsinos are also quite
important and
even the Bino is non-negligible. Compare $P_{222}$, $P_{233}$,
$P_{244}$, $P_{212}$, $P_{213}$, $P_{214}$ and $P_{211}$.
A similar situation appears also for $\tchi_4$, but with a smaller
Bino and a larger Higgsino contributions.

Finally for the third neutralino, we again have
$\tchi_3 \sim  (\tilde H_1^0 + \tilde H_2^0)/\sqrt{2} $ with
$\eta_3=-1$

\end{itemize}

Before concluding the discussion of the $P_{j\alpha\beta}$ matrix
elements in the various scenarios of \cite{Ellis-bench} on the
basis of Table 2,  we should emphasize that they have been
calculated at the rather large scale $Q$ given in the second
column of Table 1. These numbers give then, only a rough
indication of the situation.
 In order to get the corresponding matrix
elements affecting the actual production of a specific neutralino,
 we should of course determine  $P_{j\alpha\beta}$
by substituting in (\ref{Projector-c}, \ref{tProjector})
the values of  $M_1,~ M_2$ and $\mu$ at the appropriate scale of
the physical situation involved.

\vspace{0.5cm}
We next turn to the discussion of the implication of these results
to the $e^- e^+\to \tchi_i \tchi_j$ cross section
presented in (\ref{chiij-cross}-\ref{ZseR-int}).
As an example, we concentrate to the
$\tchi_1 \tchi_2$ production in the scenarios A,C,D,G,I,L.
Since then, $\tchi_1$ is almost a pure Bino, and
$\tchi_2$ a pure Wino, the dominant  contribution  arises
from terms involving a $P_{111}P_{222}$ product,  which can only
come  from the $\tilde e_L$-square contribution in
(\ref{seL-square}). Correspondingly, the $\tchi_2\tchi_2$
production mainly arises from $\tilde e_L$ exchanged
described by   (\ref{seL-square}).

In the same scenarios, the $\tchi_1 \tchi_3$
or $\tchi_1 \tchi_4$ productions are   strongly suppressed,
since $\tchi_3$ and $\tchi_4$ are almost pure Higgsinos, and there
is no Feynman diagram inducing a Bino-Higgsino production.
Because of this, in  (\ref{Z-square}-\ref{ZseR-int}) there is
never any contribution involving products of the form
$P_{111}P_{333}$, $P_{111}P_{344}$,  $P_{111}P_{334}$
or $P_{111}P_{433}$, $P_{111}P_{434}$,  $P_{111}P_{444}$.\par

\section{Summary}

Restricting to real values of  the $(M_1,~M_2)$ SUSY breaking
parameters and of the SUSY scalar sector parameter $\mu$; we have
constructed  explicit expressions  for the
neutralino projector matrices and the
signed neutralino mass eigenvalues.

We have then  shown that these
quantities are sufficient for expressing any physical observable
related to neutralinos. In fact, the cross section for
any process involving a $\tchi_j$ neutralino
in the final state, will just be proportional
to matrix elements of  its projector $P_j$, and may also
depend on its signed mass
 $\tilde m_{\tchi_j}=\eta_j m_{\tchi_j} $. These quantities
contain therefore, all physically relevant information concerning
the specific neutralino.

If more than one external neutralinos are
involved in a process, then  in a physical observable,
  each neutralino will be described by
its own projector matrix elements and signed mass.

Moreover, the projectors and signed neutralino
masses,  are   also  sufficient to
describe   any virtual neutralino exchange
contribution in MSSM, to any order in perturbation theory.

These facts, together with their easy
construction from the parameters in the SUSY Lagrangian, should
make the  use of projectors a rewarding approach for calculating
neutralino processes.

To convince the reader for that, we have
presented the $e^-e^+ \to \tchi_i \tchi_j$ differential cross
section in terms of the neutralino projector matrix elements.
As an orientation on their expected magnitude, we have also
given  their values at the scale
$Q=\sqrt{m_{\stop_1} m_{\stop_2}}$, for all SUGRA
benchmark scenarios of \cite{Ellis-bench}, which  can accommodate
that at least some of the   neutralinos will be within the reach of
LHC or the future Colliders.

On the basis of these we could
certainly claim  that the use of the projectors strongly
emphasizes the physical origin of the various terms in neutralino
related observables.

 Finally a word of caution should be added.
 The  claim that the contribution of any neutralino can be
  completely described by the corresponding
  projector matrix and signed mass, is only made
  for real $(M_1, ~M_2,~\mu)$-parameters  in the MSSM Lagrangian.
 For complex  $(M_1, ~M_2,~\mu)$, additional information
 is  needed. The investigation of this problem
 is beyond our present scope.

\end{document}